# Landau Theory Treatment of the Wurtzite-based Heterovalent Ternary Semiconductors

Paul C. Quayle

1) Department of Electrical and Computer Engineering, Michigan State University, 428 S. Shaw Lane, East Lansing, Michigan, 48824-1226, USA

2) Center for Coatings and Diamond Technologies, Fraunhofer USA, Inc., 1449 Engineering Research Court, East Lansing, Michigan, 48824-1226, USA

3) Kyma Technologies Inc., 8829 Midway West Rd, Raleigh, NC 27617, USA

Email: quaylepa@msu.edu

Phone: 517-355-5066

Fax: 517-353-1980

Abstract:

Characterizing the crystalline disorder properties of the heterovalent ternary semiconductors continues to challenge solid-state theory. Here, a Landau theory is developed for the wurtzite-based ternary semiconductor $ZnSnN_2$. It is shown that the symmetry properties of two nearly co-stable phases, with space groups *Pmc*$2_1$ and *Pbn*$2_1$, infer that a reconstructive phase transition is the source of crystal structure disorder via a mixture of the phases. The site exchange defect, which consists of two adjacent antisite defects, is identified as the nucleation mechanism of the transition. A Landau potential based on the space group symmetries of the *Pmc*$2_1$ and *Pbn*$2_1$ phases is constructed from the online databases in the ISOTROPY Software Suite and this potential is consistent with a system that undergoes a paraelectric-antiferroelectric phase transition. It is hypothesized the low temperature, *Pbn*$2_1$ phase is antiferroelectric within the c-axis basal plane. The dipole arrangements within the *Pbn*$2_1$ basal plane yield a nonpolar spontaneous polarization and the electrical susceptibility derived from the Landau potential exhibits a singularity at the Néel temperature characteristic of antiferroelectric behavior. These results inform the study of disorder in the broad class heterovalent ternary semiconductors, including those based on the zincblende structure, and opens the door to the application of the ternaries in new technology spaces.

1. <u>Introduction</u>

The heterovalent ternary semiconductors with stoichiometry I-III-VI$_2$ and II-IV-V$_2$ have more complex atomic arrangements than their binary parent semiconductors, the II-VIs and III-Vs, due to the added degree of freedom in the two-cation sublattices. Characterizing and manipulating this complexity challenges the capabilities of both theorists and experimentalists. Solid-state phase transitions are detected in many ternaries (Berger & Prochukhan, 1969; Shay & Wernick, 1975; Zunger, 1987). Mixed phase lattices, point defects, defect complexes, and line and planar defects further enrich this picture (Zhang *et al.,* 1998; Álvarez-García *et al.,* 2005; Oikkonen *et al.,* 2014; Abou-Ras *et al.,* 2016). The complexity affects and drives the performance of ternary semiconductors which enable important technologies including nonlinear optics (Petrov, 2012), thermoelectric generators (Ritz & Peterson, 2004; Cook *et al.,* 2007; Ma *et al.,* 2013), and thin film photovoltaics (Jackson *et al.,* 2011; Siebentritt, 2017). Density functional theory (DFT) (Wei *et al.,* 1999; Lyu *et al.,* 2019), and Monte Carlo have been used to characterize this complexity (Wei *et al.,* 1992; Ludwig *et al.,* 2011; Ma *et al.,* 2014; Lany *et al.,* 2017).

Landau theory (LT) has been used to study the ternaries as well (McConnell, 1978; Folmer & Franzen, 1984). LT is based on the principle that a solid-state phase transition is accompanied by a change in crystal structure symmetry. Identifying the symmetries of the phases in the transition provides information on the temperature-dependent and pressure-dependent atomic arrangements of the crystal. LT is particularly useful because it can provide information about crystalline materials that contain structure that is not periodic, making it well-suited for the analysis of materials that exhibit domain structure and nanostructure (Janovec, 1989; Müller, 2017). First introduced in the 1930s (Landau, 1937; Landau & Lifshitz, 1959; Cowley, 1980), LT became accessible to nonspecialists in 1983 when space group symmetry relations were consolidated in the International Tables for Crystallography. Since then, complete databases have been made available online along with a wealth of computational tools designed to enable Landau analysis.

Landau theory played a central role in overcoming challenges in the development of the chalcogen-based ternary semiconductors, which compose the commercial solar cell material Cu(In,Ga)Se$_2$. In its early stage, the effort to synthesize device-quality CuInS$_2$ suffered from severe cracking and nanoprecipitate formation due to a solid-state phase transition (Binsma *et al.,* 1980; Arsene *et al.,* 1996; Mullan *et al.,* 1997). LT was used to characterize the phase transition and helped establish the use of non-equilibrium vapor-phase growth methods and growth temperatures significantly lower than the Curie temperature ($T_C$) to avoid secondary phase precipitates (Folmer & Franzen, 1984; Su *et al.,* 2000; Abou-Ras *et al.,* 2016).

While growing at temperatures well below $T_C$ mitigated the detrimental effects of the phase transition in chalcogen-based ternary functional materials, in other ternaries, the disorder generated by the phase transition is advantageous. ZnSnP$_2$ is a candidate photovoltaic semiconductor with an intrinsic band gap of approximately 1.7 eV and $T_C$ of 720 °C (Nakatsuka *et al.,* 2017). Synthesis of ZnSnP$_2$ has been done using both equilibrium solution methods at temperatures above $T_C$, and far-from-equilibrium molecular beam epitaxy at temperatures well below $T_C$. In both cases, tuning the cool down rate from the growth temperature determines the degree of crystal structure disorder. Increasing the cool down rate decreases the band gap of the material. Thus, ZnSnP$_2$ is an adjustable band gap photovoltaic material (Ryan *et al.,* 1987; Nakatsuka & Nose, 2017).

Understanding and controlling disorder in ternaries like ZnSnP$_2$ requires a precise characterization of the phases above and below the phase transition point. The identification of the two phases and of the nature of the atomic scale structure that generates lattice disorder are the main subjects of this paper. This analysis is called for because the validity of the historically predominant model of lattice disorder in the ternaries has been questioned by both experiment and theory.

The predominant model holds that the high symmetry phase above the $T_C$ is characterized by an entropically random distribution of cations. This theory was first put forward by Buerger (1934) and served as the basis of the first LT analyses of the chalcogen-based ternaries (McConnell, 1978; Folmer & Franzen, 1984). The random disorder model states that thermal energy randomizes the positions of the bivalent cation sublattice at the phase transition, breaks the symmetry of the low symmetry, chalcopyrite phase and yields an isotropic cation sublattice with symmetry that is equivalent to the that of the sphalerite ZnS structure. Evidence for the random disorder model comes primarily from XRD (Shay & Wernick, 1975). Measurements of the zinc-blende-based chalcopyrite ternaries taken near and above the Curie temperature only show peaks that are consistent with the sphalerite structure.

X-ray diffraction however, while useful for characterizing the macroscropic symmetry of the crystal, gives no information about the local environment of the atoms, since the detected signal is the average over the coherence length of x-rays. Studies using other methods yield data that conflict with the random disorder model. For example, band gap measurements of $ZnSnP_2$ show a decrease of 18% and 22% when comparing ordered samples to disordered (Ryan *et al.*, 1987; St-Jean *et al.*, 2010). DFT calculations based on the random model predict much larger decreases in band gap for the two cases of 56% and 76% (Scanlon & Walsh, 2012; Ma *et al.*, 2014).

Ma *et al.* (2014) concluded based on DFT that a random distribution is not possible under equilibrium conditions, a position that has been supported (Skachkov *et al.*, 2016; Lany *et al.*, 2017). The random arrangement of atoms cannot exist under equilibrium conditions because it generates too many instances in which the octet rule is violated. In the lowest energy crystal structures of ternaries such as $ZnSnP_2$, the lattice is populated according to the octet rule; every group V anion is bonded to two group

II and two group IV cations, forming $Zn_2Sn_2$ tetrahedra. On a randomized cation sublattice, there are a statistical number of tetrahedra in which a group V anion is bonded to three group II and one group IV cations ($Zn_3Sn_1$), or vice versa, and four group II and zero group IV cations ($Zn_4Sn_0$), or vice versa. The $Zn_3Sn_1$ and $Zn_1Sn_3$ tetrahedra and the $Zn_4Sn_0$ and $Zn_0Sn_4$ tetrahedra, especially, violate the octet rule and have high formation energies. Ma *et al.* (2014) predict based on Monte Carlo simulations that, as the temperature of $ZnSnP_2$ is increased, thermal energy generates an increasing probability that $Zn_3Sn_1$ and $Zn_1Sn_3$ will form; above a phase transition at ~1100 K, this probability approaches 20%. Their results also predict that, even at 20000 K, the probability of $Zn_4Sn_0$ and $Zn_0Sn_4$ tetrahedra formation is lower than required for a randomized lattice.

Ryan *et al.* (1987) presented an alternative to the random disorder model. The study included a comparison of nuclear magnetic resonance spectra taken from ordered and disordered $ZnSnP_2$. A pair of peaks appeared in the disordered samples which they assigned to the $Zn_3Sn_1$ and $Zn_1Sn_3$ tetrahedra. The data did not show any additional peaks that could be assigned to the $Zn_4Sn_0$ and $Zn_0Sn_4$ tetrahedra however leading them to the conclusion that those tetrahedra are not present on the lattice, and that the lattice disorder is not random. In place of the random disorder model, Ryan *et al.* (1987) proposed that the disorder was caused by a scattering of domains embedded within the chalcopyrite structure. They hypothesized that these domains consisted of arrangements of neighboring $Zn_3Sn_1$ and $Zn_1Sn_3$ tetrahedra generated by an exchange in position of adjacent Zn and Sn atoms, and that these site exchange defects (SEDs) create mutually compensating acceptor-donor pairs.

The model proposed by Ryan *et al.* (1987) has received interest lately in investigations of the wurtzite-based ternary $ZnSnN_2$. $ZnSnN_2$ has emerged as a promising photovoltaic material composed of earth-abundant elements. Like $ZnSnP_2$, a range of band gaps have been reported for $ZnSnN_2$, spanning 1.4 eV to 2.0 eV (Martinez *et al.*, 2017). Raman measurements reported by Quayle *et al.* (2015) show a glass-like spectrum consistent with a nonperiodic lattice. All XRD measurements of ZnSnN$_2$ to date show a macroscopically disordered spectrum consistent with wurtzite (Lyu *et al.,* 2019).

Recent studies of ZnSnN$_2$ have incorporated both the SED and domain mixtures into models of ZnSnN$_2$. Lany *et al.* (2017) predict based on DFT that the SED costs approximately 0.04 eV/pair relative to the most stable structure. By including Zn$_3$Sn$_1$, Zn$_1$Sn$_3$, Zn$_4$Sn$_0$, and Zn$_0$Sn$_4$ tetrahedra in Monte Carlo simulations of ZnSnN$_2$, the authors showed results like those reported by Ma *et al.* (2014) for ZnSnP$_2$. At elevated temperatures between ~1500-2000 °C, the density of Zn$_3$Sn$_1$ and Zn$_1$Sn$_3$ tetrahedra increases to a concentration of around 10%. They reported that the density of Zn$_3$Sn$_1$ and Zn$_1$Sn$_3$ tetrahedra has only a moderate effect on the band gap; lowering it by ~0.3 eV. Recently, Makin *et al.* (2019) reported that the band gaps of ZnSnN$_2$ and MgSnN$_2$ can be decreased via disorder tuning from 1.98 eV to 1.12 eV and 3.43 eV to 1.87 eV, respectively.

First principles studies agree that the most stable crystal structure of ZnSnN$_2$ is orthorhombic *Pbn*2$_1$ (*Pna*2$_1$ in the standard setting), and that a second competing phase with space group *Pmc*2$_1$ is slightly less favorable (Fig. 2) (Lyu *et al.,* 2019; Martinez *et al.,* Lahourcade *et al.,* 2013). (Note that the non-standard setting is used for the *Pbn*2$_1$ phase so that its **a** and **b** lattice parameters and the **a** and **b** lattice parameters of the *Pmc*2$_1$ phase are aligned in the same direction.) Quayle *et al.* (2015) predict that the formation energy of the *Pmc*2$_1$ phase is higher than *Pbn2$_1$* by only 0.011 ± 0.003 eV, and hypothesize that the *Pbn*2$_1$ and *Pmc*2$_1$ phases form a mixed lattice that consists of *Pbn*2$_1$ or *Pmc*2$_1$ basal planes stacked along the polar c-axis, similar to SiC polytypes. XRD simulations of the mixed phase crystals show that this type of basal plane disorder 'washes out' Bragg reflections unique to the *Pbn*2$_1$ and *Pmc*2$_1$ crystal structures, yielding a wurtzite-like spectrum. These results provide an explanation for the observations of a wurtzite-like XRD spectrum that does not assume a randomly disordered lattice.

In this paper, we investigate a Landau theory of ZnSnN$_2$ assuming that a reconstructive phase transition takes place between the low temperature $Pbn2_1$ phase and the high temperature phase $Pmc2_1$ phase. The SED is proposed to be the mechanism of the transition. Skachkov *et al.* (2016) and Adamski *et al.* (2017) determined that the SED is one of the lowest formation energy defect in ZnSnN$_2$ and its sister compound ZnGeN$_2$. Skachkov *et al.* (2016) also determined that there is an energetic benefit to the clustering of SEDs in these materials; the formation energy of two neighboring SEDs is lower than two isolated SEDs. Here, we see that clusters of SEDs transform the $Pmc2_1$ crystal structure into the $Pbn2_1$ crystal structure. An analysis of the group-subgroup relation of the two phases shows that the phase relation is not direct and there are two intermediate phases that mediate the phase transition. The order parameter of the phase transition is obtained from the online databases along with the Landau potential. A solution to the Landau potential is given based on the analyses of similar free energy equations in the literature.

The model presented here has broad implications for all heterovalent ternary semiconductors. Crystal structure disorder and phase transitions have been widely reported in the zincblende-based ternaries, however, until Ma *et al.* (2014), it was widely accepted that the disordered phase above $T_C$ is caused by a random high temperature phase. In the zincblende-based case, the situation is analogous to the wurtzite-based ternaries; there are exactly two atomic arrangements that satisfy the octet rule, the chalcopyrite ($I\bar{4}2d$, #122) and CuAu (P$\bar{4}m2$, #115) phases. High resolution electron diffraction and electron microscopy clearly resolve a mixture of the chalcopyrite and CuAu phases in CuInS$_2$ and CuInSe$_2$ (Álvarez-García *et al.,* 2005; Su *et al.,* 2000; Su & Wei, 1999; Metzner *et al.,* 2000; Stanbery *et al.,* 2002). Furthermore, it is argued here that the wurtzite-based ternaries are antiferroelectric in the c-plane. If confirmed experimentally, the introduction of a new class of antiferroelectric semiconductors is of both fundamental and practical interest.

The remainder of the paper is organized as follows. First, we illustrate how the *Pbn*2$_1$ and *Pmc*2$_1$ phases intermix via the clustering of SEDs. Next, we analyse the group-subgroup relations between the *Pmc*2$_1$, *Pbn*2$_1$ and intermediate phases. Following that, we use the group-subgroup relations to identify the order parameter and Landau potential of the system. The Landau potential is then solved under the assumption of strong coupling between the phases. A brief discussion of the prospect that the wurtzite-based ternaries are antiferroelectric materials is given before a summary concludes the paper.

2. Illustration of phase mixing

Building off the model proposed by Ryan *et al.* (1987) for ZnSnP$_2$, we investigate the atomic formations generated by clusters of SEDs within the ZnSnN$_2$ lattice. In this section, we see that the *Pmc*2$_1$ crystal structure can transform into the *Pbn*2$_1$ structure via the intermediate phases with space groups *Pmn*2$_1$ and *Pbc*2$_1$. (*Pca*2$_1$ in the standard setting).

Figure 1 shows the primitive unit cells of the four phases of ZnSnN$_2$. The two phases on the left are the two lowest energy phases; they correspond to the two ways that the lattice can be populated so that each N-centered tetrahedron is type -Zn$_2$Sn$_2$ and the octet rule is satisfied. The two phases on the right do not satisfy the octet rule. Each tetrahedron is type -Zn$_3$Sn$_1$ or -Zn$_1$Sn$_3$.

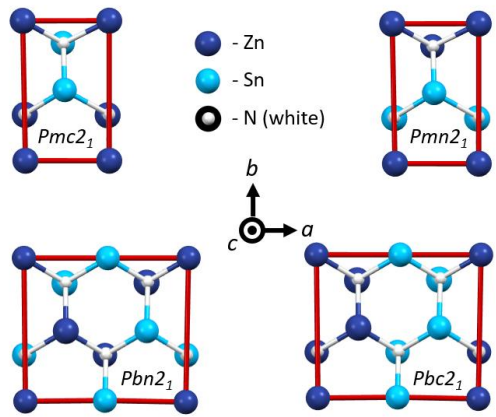

Figure 1. The phases of ZnSnN$_2$ involved in the transition from *Pmc*2$_1$ to *Pbn*2$_1$. Primitive unit cells are outlined in red.

Figure 2 shows a domain of the *Pbn2₁* phase embedded within the *Pmc2₁* background. The unit cell at the top of Fig. 2 highlights the 8-atom *Pmc2₁* primitive unit cell. The unit cell at the bottom highlights a 32-atom *Pbn2₁* unit cell. There are two types of SED patterns that lead to the *Pbn2₁* phase. One type of pattern yields an atomic arrangement consistent with the *Pbc2₁* space group, the second yields a *Pmn2₁* unit cell. The *Pbn2₁* atomic arrangement is generated when the two SED patterns coincide.

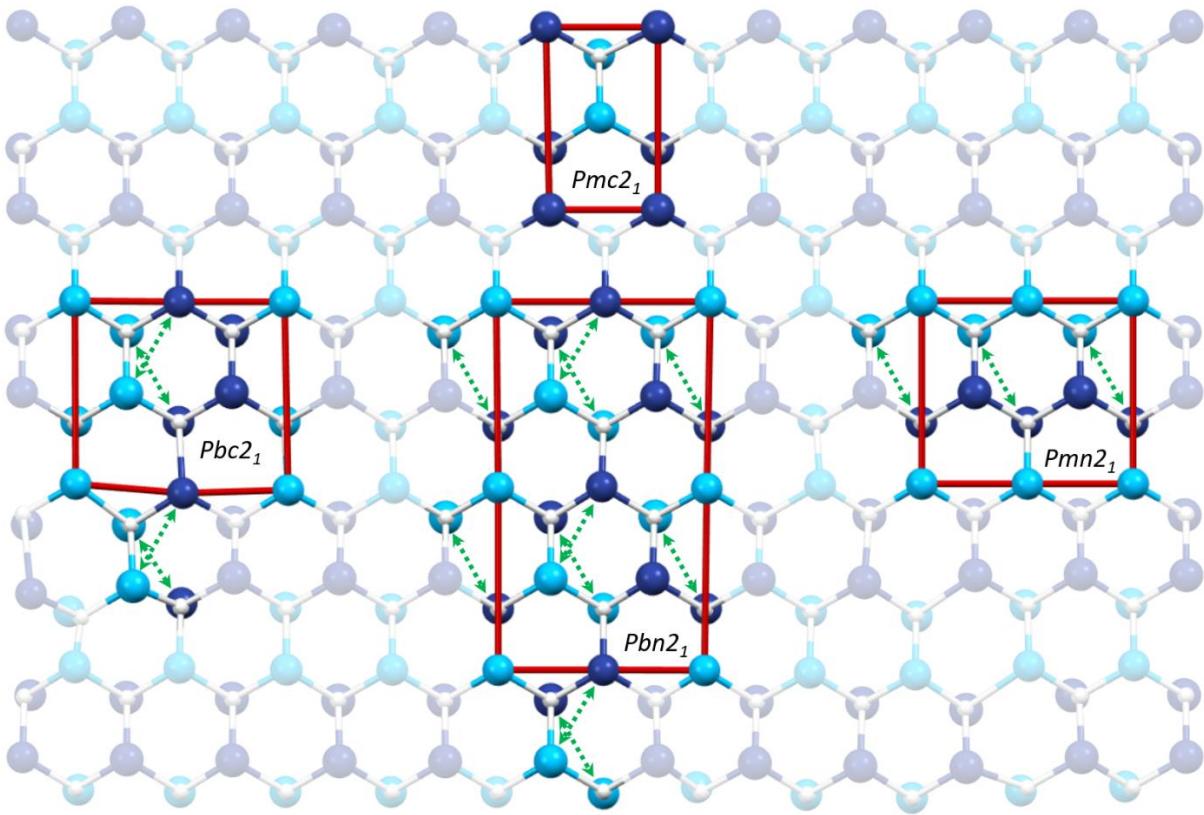

Figure 2. Illustration of the subgroup phases atomic arrangements embedded within the *Pmc2₁* crystal structure. The transformation of *Pmc2₁* phase into the subgroup phases is induced by the SEDs, indicated by the green arrows. Unit cells are outlined in red.

3. Group-subgroup relation of the ZnSnN$_2$ phases

A group-subgroup analysis of ZnSnN$_2$ is based on the similarities and differences in symmetry of the *Pmc*2$_1$ phase and *Pbn*2$_1$ phase. The group elements of a space group are the symmetry operations - the rotations and/or translations - that transform the crystal structure back into itself. A group that results from the removal of one or more of the symmetry operations is a subgroup. The program SUBGROUPGRAPH on the Bilbao Crystallographic Server gives that the *Pbn*2$_1$ space group is a subgroup of *Pmc*2$_1$ with the group-subgroup index ($i$) of 4 (Ivantchev *et al.*, 2000), and that the group-subgroup relation is indirect. There are three possible chains linking the two groups, *Pmc*2$_1$ > (*Pbc*2$_1$, *Cmc*2$_1$, *Pmn*2$_1$) > *Pbn*2$_1$ (Fig. 3). For the intermediate relations, *Pmc*2$_1$ > (*Pbc*2$_1$, *Cmc*2$_1$, *Pmn*2$_1$) and (*Pbc*2$_1$, *Cmc*2$_1$, *Pmn*2$_1$) > *Pbn*2$_1$, the subgroup indices are $i = 2$.

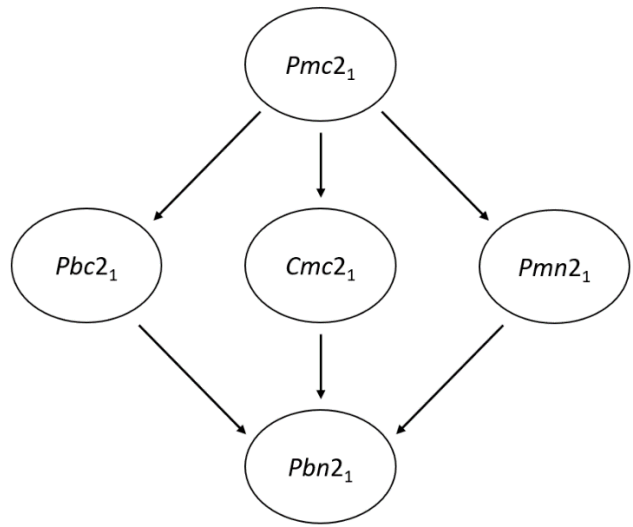

Figure 3. Diagram of the *Pmc*2$_1$-*Pbn*2$_1$ group-subgroup phase relation.

Only the *Pmn*2$_1$ and *Pbc*2$_1$ phases are valid intermediate phases. The *Cmc*2$_1$ phase is not compatible with the orthorhombic crystal structure. To demonstrate this incompatibility, we examine further the subgroup indices.

The subgroup index relates the number of symmetry operations in the space group of the group phase to those of the subgroup phase,

$$i = \frac{Z(subgroup)}{Z(group)} \cdot \left|\frac{P(group)}{P(subgroup)}\right| \qquad (1)$$

where $Z$ is the number of formula units per unit cell, and $|P|$ is the order of the point group. Each of the space groups has the same point group mm2, meaning that the second term in (1) can be dropped, and that the subgroup index is simply the ratio of the number of formula units in the two phases.

Equation 1 tells us that the unit cell of the $Pbn2_1$ structure ($i = 4$) included in the phase transition contains 32 atoms, since the $Pmc2_1$ structure has an 8-atom primitive unit cell, or two formula units. The unit cells of the intermediate phases ($i = 2$) each contain 16 atoms.

The atomic arrangements of the 16-atom orthorhombic unit cells are subject to constraints at the boundaries to maintain periodicity. The constraints are: 1) The cations at the vertices of the orthorhombic unit cell must the same and 2) the cations positioned on opposite faces of the unit cells must be the same. There are 35 possible unit cells that satisfy these constraints. Each of the allowed unit cells is listed in Appendix A along with its space group, which was determined using the ISOTROPY Software Suite program FINDSYM (Stokes & Hatch, 2005). The results show that the $Cmc2_1$ space group is incompatible with the 16-atom orthorhombic ZnSnN$_2$ lattice.

4. Determining the order parameter and Landau potential

A solid-state phase transition is characterized by a shift in the positions of atoms on the lattice. The atomic displacements break the symmetry of the high symmetry group phase, yielding a crystal structure with the decreased symmetry of the subgroup phase. In a single crystal, the atomic displacements are periodic, and they can be collectively associated with a normal mode of the system.

The order parameter of the phase transition is often assigned to the normal mode wavevector, although it can be insightful to assign it to a function of the wavevector. The distortion gives rise to a polarization vector in ferroelectric crystals, for example. The amplitude of the displacements from the positions in the high symmetry phase decreases as the temperature is increased towards the transition point, and the order parameter approaches zero. Accordingly, we can express the free energy of the system as a function of the order parameter and Taylor expand around $T_C$ for small values of the order parameter. The energy of the system, written as a function of the order parameter, is the Landau potential.

Each group-subgroup relation is associated with one or more order parameters and these order parameters were calculated for every group-subgroup possibility for all 230 space groups by Hatch & Stokes (1988). The complete listing of their results is available in the online database, COPL (Hatch & Stokes, 2002; Hatch & Stokes, 2002), and the results for our case are listed in Table 1.

| k-vectors | Irreps and Order Parameters | Isotropy Subgroup |
|---|---|---|
| GM: **(0,0,0)** | GM$_1$: (a) | Pmc2$_1$ |
| X: **(1/2,0,0)** | X$_2$: (a) | Pmn2$_1$ |
| Y: **(0,1/2,0)** | Y$_2$: (a) | Pbc2$_1$ |

Table 1. Group-subgroup data for the *Pmc*2$_1$ → *Pbn*2$_1$ transition.

Since the *Pbn*2$_1$ phase is generated by the overlap of the atomic displacements associated with both the *Pbc*2$_1$ and *Pmn*2$_1$ phases (Fig. 2), both order parameters are required for the phase transition and the Landau potential which expresses the free energy of the transition is based on the coupled order parameter. We find the Landau potential from the ISS program INVARIANT (Stokes & Hatch, 2003),

$$\Phi = \Phi_0 + \frac{A}{2}a^2 + \frac{B}{2}b^2 + \frac{C}{4}a^4 - \frac{D}{2}a^2b^2 + \frac{E}{4}b^4 + \frac{F}{6}a^6 - \frac{G}{2}a^4b^2 - \frac{H}{2}a^2b^4 + \frac{I}{6}b^6 \qquad (2)$$

where $a$ is one order parameter and $b$ is the other. The coefficient $A$ is assumed to be temperature dependent,

$$A = \frac{1}{\Gamma}(T - T_a). \qquad (3)$$

All other constants are constant and positive. The sixth order terms must add up to be positive so that the free energy is positive at the extremes. The sign of the coupling terms is critical in determining the nature of the system. Choosing the coupling terms to be negative allows for the description of triggered phase transitions. In the next section, it will be argued that ZnSnN$_2$ necessarily undergoes a triggered phase transition.

5. Solution to the Landau potential

Holakovsky (1973) established that the necessary conditions for a triggered ferroelectric transition is that the lowest order coupling term is of the form $a^2b^2$ and that the sign of the term is negative. He considered materials that transition from paraelectric to ferroelectric, and that have both a primary and secondary order parameter. The secondary order parameter is assigned to a polarization vector, and it is shown that a phase transition driven by the primary order parameter generates a second, ferroelectric phase transition, as a result of the coupling. The behavior of the system is determined by the strength of the coupling terms. Under sufficiently strong coupling conditions, a triggered ferroelectric transition will occur in which both phase transitions occur simultaneously. This condition for strong coupling in ZnSnN$_2$ is determined in Appendix B.

The analysis of Holakovsky (1973) built off the work of Levanyuk & Sannikov (1969), which analyzed systems with different types of coupling terms. Their work established that systems with $a^2b^2$ coupling terms are usually associated with antiferroelectric materials. Accordingly, we look for characteristics of antiferroelectric polarization in ZnSnN$_2$.

The crystal structure of the wurtzite-based ternaries suggests that ZnSnN$_2$ is antiferroelectric within the c-axis basal plane. In the orthorhombic II-IV-V$_2$ compounds, the group IV atoms each transfer an electron to the group II atoms, so that the group II atoms are negatively charged and the group IV atoms are positively charged (Ma *et al.,* 2014; Skachkov *et al.,* 2016). As a result, the basal plane consists of an array of dipoles. Adding up the dipoles yields a series of equal but oppositely pointing net polarization vectors (Fig. 4). In the *Pmc*2$_1$ basal plane, the polarization vectors cancel out completely making it paraelectric.

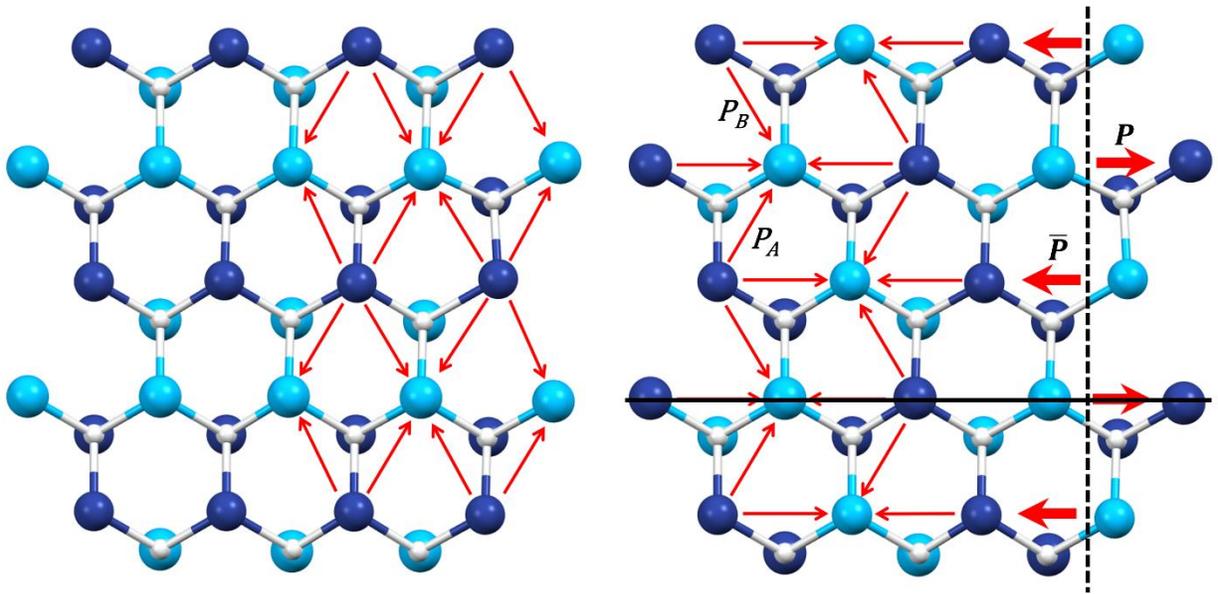

Figure. 4. C-plane bonding characteristics of the paraelectric *Pmc*2$_1$ phase (left) and antiferroelectric *Pbn*2$_1$ phase (right). The thin arrows indicate the directions of the polarization vectors. The larger arrows indicate the directions of the net polarizations vectors. The dashed line is along a glide-mirror plane. The solid line is along a mirror plane of the two-dimensional, cation sublattice plane.

Thus, we hypothesize that a solid-state phase transition in ZnSnN$_2$ from *Pmc*2$_1$ to *Pbn*2$_1$ is paraelectric-antiferroelectric. Furthermore, the coupling of the order parameters is not merely strong, it is *locked*. In the same way that the secondary order parameter is assigned to the polarization vector in triggered ferroelectric transition, both order parameters can be assigned to the two polarization vectors in the antiferroelectric $Pbn2_1$ phase. The net polarization vectors are generated by the SEDs, and the neighboring SEDs that form a $Pbn2_1$ domain within $Pmc2_1$ generate two oppositely polarized vectors. The phase transition cannot proceed further than a single SED if both order parameters are not triggered. This amounts to a simultaneous transition of both order parameters; the system is inherently strongly coupled and both order parameters are primary.

Following Levanyuk & Sannikov (1969), we analyze the system under the assumption of antiferroelectric behavior and use the polarization vectors,

$$P_A = \tfrac{1}{2}(a + b) \tag{4}$$

$$P_B = \tfrac{1}{2}(a - b) \tag{5}$$

The $P_A$ and $P_B$ polarization vectors are represented in Fig. 4 and they correspond to the dipoles that summate to finite net polarization vector, $P = P_A + P_B$. The antiferroelectric polarization arises from the glide mirror plane that runs along the y-axis points; there is a net polarization vector $\bar{P}$ pointing in the opposite direction to and shifted along the y-axis from $P$, which is equivalent by symmetry.

Using (4) and (5), we rewrite (2) in terms of the polarization vectors,

$$\Phi = \Phi_0 + \alpha\left(P_a^{\ 2} + P_b^{\ 2}\right) - \gamma P_a^{\ 2} P_b^{\ 2} + \beta\left(P_a^{\ 4} + P_b^{\ 4}\right) - \delta\left(P_a^{\ 4} P_b^2 + P_a^{\ 2} P_b^{\ 4}\right) + \varepsilon\left(P_a^{\ 6} + P_b^{\ 6}\right) \tag{6}$$

where, $\alpha = A, \beta = C/2 - D/2, \gamma = -(3C + D), \delta = -(5F + G), \varepsilon = F/3 - G$. In writing (6), we make use of the symmetry due to the mirror plane in the cationic basal plane that makes $A = B, C = E, F = I$, and $G = H$.

The free energy expressed in (6) is similar to the one described by Kittel (1950), and we can evaluate it using his methods.

In the antiferroelectric regime under no applied electric field, the spontaneous polarization is $P_{s,a} = -P_{s,b}$. Minimizing (6) yields,

$$\alpha + P_{s,a}^2(2\beta - \gamma) + 3P_{s,a}^4(\varepsilon - \delta) = 0 \tag{7}$$

At the transition point, the local minima expressed by (7) is equal to the potential at $P_{s,a} = P_{s,b} = 0$,

$$2\alpha + P_{s,a}^2(2\beta - \gamma) + 2P_{s,a}^4(\varepsilon - \delta) = 0 \tag{8}$$

Using (7) and (8) we find,

$$P_{s,a}^2 = -\frac{4\alpha}{2\beta - \gamma} \tag{9}$$

$$P_{s,a}^4 = \frac{\alpha}{\varepsilon - \delta} \tag{10}$$

If we apply a small electric field $\Delta E$, the net polarization is $\Delta P = P_a + P_b$, where $P_a \cong -P_b$. Taking the sum of $\frac{\partial \Phi_{AFE}}{\partial P_a} = \Delta E$ and $\frac{\partial \Phi_{AFE}}{\partial P_b} = \Delta E$ yields,

$$2\Delta E = 2\alpha(P_a + P_b) + 2(2\beta - \gamma)(P_a^3 + P_b^3) + 6(\varepsilon - \delta)(P_a^5 + P_b^5) \tag{11}$$

Since, $P_a = P_{s,a} + \frac{1}{2}\Delta P$ and $P_b = P_{s,b} + \frac{1}{2}\Delta P$, the susceptibility just below the transition temperature, now the Néel temperature ($T_N$), is

$$\chi = \frac{\Delta P}{\Delta E} = \frac{1}{4\alpha} \tag{12}$$

Above $T_N$, the higher order terms in (6) can be neglected and

$$\chi = \frac{1}{\alpha} \tag{13}$$

By setting $\Gamma = 1$, we can plot the susceptibility (Fig. 5) and see the singularity at $T_N$.

The condition for $T_N$ is found using (9) and (10):

$$2\beta - \gamma = -4\sqrt{\alpha(\varepsilon - \delta)} \qquad (14)$$

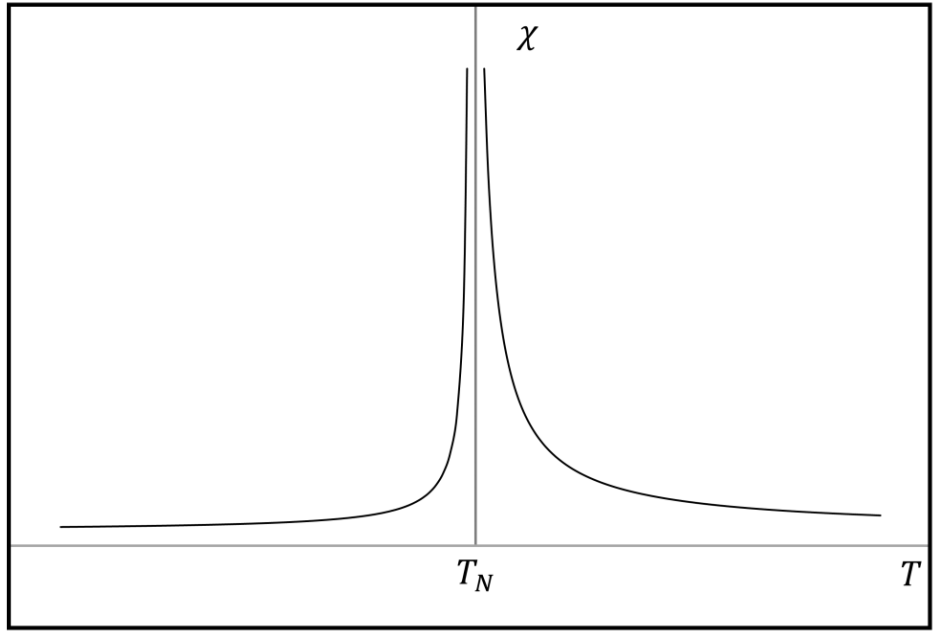

Figure 5. Temperature dependence of the electrical susceptibility for a first order, antiferroelectric phase transition.

6. Discussion

The temperature dependence of the electrical susceptibility plotted in Figure 5 displays the anomaly at the Néel temperature observed in the dielectric response of antiferroelectric materials such as PbZrO$_3$ (Whatmore & Glazer, 1979; Fthenakis & Ponomareva, 2017; Shirane *et al.*, 1951; Roberts, 1949; Liu & Dkhil, 2011, Tagantsev *et al.*, 2013). Whether ZnSnN$_2$ is a true antiferroelectric material depends on additional criteria however (Rabe, 2013).

The non-polar, antiferroelectric polarization in ZnSnN$_2$ arises from the structural distortion generated by the SED formations which transform the paraelectric *Pmc*2$_1$ phase. To observe the P vs. E double hysteresis loop characteristic of antiferroelectrics, the material must exhibit an additional ferroelectric distortion from the paraelectric phase when subject to an external electric field (Bennett *et al.*, 2013, Tolédano & Guennou, 2016). It is reasonable to assume that this distortion will be present. In Fig. 6 we again look at the dipole arrangement of the paraelectric *Pmc*2$_1$ phase. For simplicity, we only consider the two-dimensional, cation sublattice plane. Given the charge states of the different cations, under the application of an external field, we can expect a shift in positions of the cations due to the Coulomb force. The induced distortion will break the symmetry-based cancellation of the dipole moments, resulting in a ferroelectric polarization in the cation plane.

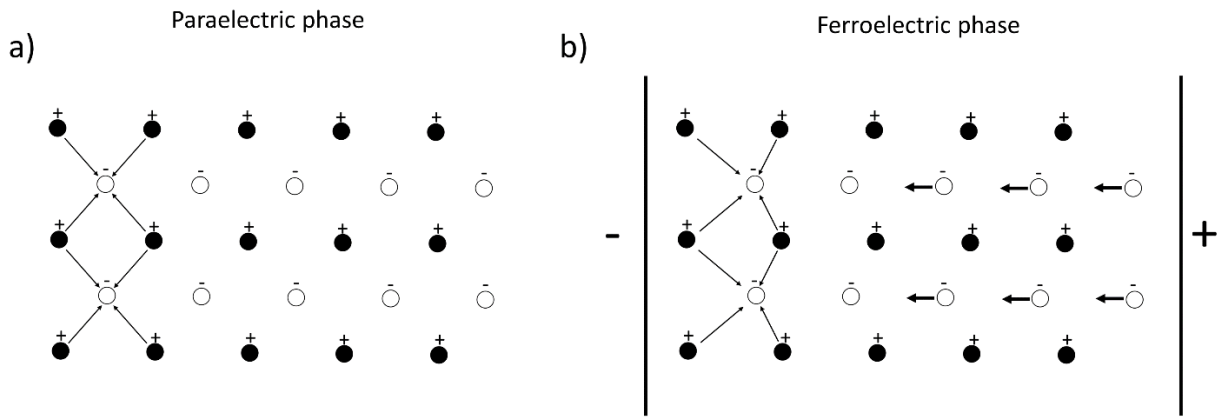

Figure 6. The high temperature *Pmc*2$_1$ phase under a) zero bias and b) an applied electric field. Only the atoms in the two-dimensional, cation sublattice plane are shown. The thin arrows represent polarization vectors generated by the charge states of the Zn and Sn atoms. The thicker arrows in b) represent the net polarization vectors generated by the electric-field induced atomic distortion.

Summary and Outlook

To summarize, based on the calculated low formation energy of the SED, and the very similar formation energies of the *Pbn*2$_1$ and *Pmc*2$_1$ crystal structures, a model for a first order, reconstructive phase transition was developed to interpret observations of structural disorder in ZnSnN$_2$. It was shown that patterns of SEDs generate the *Pbn*2$_1$ phase within the *Pmc*2$_1$ phase. A Landau potential based on two primary order parameters which activate simultaneously was analyzed. A solution to the free energy equation was derived based on the model developed by Kittel (1950) for a paraelectric-antiferroelectric transition. The electrical susceptibility expressed by Eqs. 12 and 13, and plotted in Fig. 5, is characteristic of antiferroelectric materials.

The picture described in this work is one in which thermal energy drives the formation of SEDs, which cluster to transform the atomic arrangements on the lattice. The *Pbn*2$_1$ phase is more stable than the *Pmc*2$_1$ phase at lower temperatures. As the temperature increases, the probability of SED formation increases. A single SED serves as a nucleation site for the precipitation of a domain of the *Pmc*2$_1$ phase, which then increases in size as temperature is further increased. At high temperatures, the *Pmc*2$_1$ phase is more stable than the *Pbn*2$_1$ phase and will be the dominant phase. It is likely that thermal energy will continue to generate SEDs, however, and that the high temperature phase will be dynamic, with the SED formation and annihilation varying the phase composition of the structure until the melting point.

The ternary nitrides are in the early stages of development and are little studied compared to the binary nitrides or the zincblende-based ternaries. A phase transition has not been investigated experimentally in ZnSnN$_2$. A phase transition is not the only possible cause of the disorder; kinetic barriers that inhibit atom mobility during growth are suggested to contribute to disorder in ZnGeN$_2$ (Lany *et al.,* 2017; Blanton *et al.,* 2017), and point defects, defect complexes and off-stoichiometry are proposed to contribute to disorder in the zincblende-based ternaries generally. Cation disorder is widely reported to be a major factor in the heterovalent ternaries however. The model presented here is an alternative to the one based on entropically random disorder at high temperatures, developed to describe the order-disorder transition observed in the zincblende-based ternaries.

Essential aspects of the model presented here hold in the zincblende-based ternaries, in which, there are also two phases that satisfy charge neutrality. As state in the Introduction, direct evidence of the chalcopyrite and CuAu phase mixing has been clearly observed in CuInSe$_2$, and the formation of a mixture of those two phases may be unavoidable in the CIGS system (Su *et al.,* 2000). The wurtzite-based and zincblende-based ternary systems are not completely analogous however. The space group $I\bar{4}2d$ and P$\bar{4}m2$ of the zincblende-based system are not group-subgroup related according to SUBGROUPGRAPH. Indeed, the wurtzite-based ternaries are exceptional in this regard. Typically, crystalline materials that undergo a reconstructive phase transition do not have group-subgroup relations between phases (Dmitriev & Toledano, 1996), and an order parameter cannot be defined. Thus, the ability to carry out develop a Landau theory for the wurtzite-based ternaries is noteworthy. The reason for the absence of a group-subgroup relation in the zincblende-based ternaries may be related to the fact that, while the crystal symmetry of the *Pbn*2$_1$ crystal structure results in an antiferroelectric polarization, the dipole arrangement of both the $I\bar{4}2d$ and P$\bar{4}m2$ zincblende-based phases add up to a complete cancellation and the order parameter cannot be associated with a polarization vector.

The proposal here that the wurtzite-based ternaries are antiferroelectric in the c-plane is based on the crystal structure of the *Pbn*2$_1$ phase and the form of the order parameter coupling term in the Landau potential. Validation of this model would come from observation of the double hysteresis loops in P vs. E. curve which are the signature characteristics of antiferroelectric materials (Rabe, 2013). If validated, this class of materials would hold potential for new applications. The dramatic increase in electrical susceptibility near the Néel temperature has been exploited in antiferroelectrics for capacitive energy storage technologies. The energy storage density of AFE materials is superior to that of linear and ferroelectric dielectrics (Liu *et al.,* 2018).

Acknowledgements: I thank E. Blanton for helpful discussion.

Appendix A. Intermediate phase crystal structure information

| Atomic Position | x | y | z |
|---|---|---|---|
| 1 | 0 | 0 | 0 |
| 2 | 1/2 | 0 | 0 |
| 3 | 1/4 | 1/2 | 0 |
| 4 | 3/4 | 1/2 | 0 |
| 5 | 0 | 1/3 | 1/2 |
| 6 | 1/2 | 1/3 | 1/2 |
| 7 | 1/4 | 5/6 | 1/2 |
| 8 | 3/4 | 5/6 | 1/2 |
| 9 | 0 | 0 | 3/8 |
| 10 | 1/2 | 0 | 3/8 |
| 11 | 1/4 | 1/2 | 3/8 |
| 12 | 3/4 | 1/2 | 3/8 |
| 13 | 0 | 1/3 | 7/8 |
| 14 | 1/2 | 1/3 | 7/8 |
| 15 | 1/4 | 5/6 | 7/8 |
| 16 | 3/4 | 5/6 | 7/8 |

Table 1A. Atomic positions of a 16-atom orthorhombic unit cell.

| Atom Position | | | | | | | | | | | | | | | | | Space Group |
|---|---|---|---|---|---|---|---|---|---|---|---|---|---|---|---|---|---|
| 1 | 2 | 3 | 4 | 5 | 6 | 7 | 8 | 9 | 10 | 11 | 12 | 13 | 14 | 15 | 16 | | |
| Zn | Zn | Zn | Zn | Sn | Sn | Sn | Sn | N | N | N | N | N | N | N | N | | Cm |
| Zn | Zn | Zn | Sn | Zn | Sn | Sn | Sn | N | N | N | N | N | N | N | N | | P1 |
| Zn | Zn | Zn | Sn | Sn | Zn | Sn | Sn | N | N | N | N | N | N | N | N | | P1 |
| Zn | Zn | Zn | Sn | Sn | Sn | Zn | Sn | N | N | N | N | N | N | N | N | | Pm |
| Zn | Zn | Zn | Sn | Sn | Sn | Sn | Zn | N | N | N | N | N | N | N | N | | Pm |
| Zn | Zn | Sn | Zn | Zn | Sn | Sn | Sn | N | N | N | N | N | N | N | N | | P1 |
| Zn | Zn | Sn | Zn | Sn | Zn | Sn | Sn | N | N | N | N | N | N | N | N | | P1 |
| Zn | Zn | Sn | Zn | Sn | Sn | Zn | Sn | N | N | N | N | N | N | N | N | | Pm |
| Zn | Zn | Sn | Zn | Sn | Sn | Sn | Zn | N | N | N | N | N | N | N | N | | Pm |
| Zn | Zn | Sn | Sn | Zn | Zn | Sn | Sn | N | N | N | N | N | N | N | N | | Pmc2$_1$ |
| Zn | Zn | Sn | Sn | Zn | Sn | Zn | Sn | N | N | N | N | N | N | N | N | | P1 |
| Zn | Zn | Sn | Sn | Zn | Sn | Sn | Zn | N | N | N | N | N | N | N | N | | P1 |
| Zn | Zn | Sn | Sn | Sn | Zn | Sn | Zn | N | N | N | N | N | N | N | N | | P1 |
| Zn | Zn | Sn | Sn | Sn | Zn | Zn | Sn | N | N | N | N | N | N | N | N | | P1 |
| Zn | Zn | Sn | Sn | Sn | Sn | Zn | Zn | N | N | N | N | N | N | N | N | | Pmn2$_1$ |
| Zn | Sn | Zn | Zn | Zn | Sn | Sn | Sn | N | N | N | N | N | N | N | N | | Pm |
| Zn | Sn | Zn | Zn | Sn | Zn | Sn | Sn | N | N | N | N | N | N | N | N | | Pm |
| Zn | Sn | Zn | Zn | Sn | Sn | Zn | Sn | N | N | N | N | N | N | N | N | | P1 |
| Zn | Sn | Zn | Zn | Sn | Sn | Sn | Zn | N | N | N | N | N | N | N | N | | P1 |
| Zn | Sn | Zn | Sn | Zn | Zn | Sn | Sn | N | N | N | N | N | N | N | N | | P1 |
| Zn | Sn | Zn | Sn | Zn | Sn | Zn | Sn | N | N | N | N | N | N | N | N | | Pca2$_1$ |
| Zn | Sn | Zn | Sn | Zn | Sn | Sn | Zn | N | N | N | N | N | N | N | N | | P2$_1$ |
| Zn | Sn | Zn | Sn | Sn | Zn | Sn | Zn | N | N | N | N | N | N | N | N | | Pna2$_1$ |
| Zn | Sn | Zn | Sn | Sn | Sn | Zn | Zn | N | N | N | N | N | N | N | N | | P1 |
| Zn | Sn | Zn | Sn | Sn | Zn | Zn | Sn | N | N | N | N | N | N | N | N | | P2$_1$ |
| Zn | Sn | Sn | Zn | Zn | Zn | Sn | Sn | N | N | N | N | N | N | N | N | | P1 |
| Zn | Sn | Sn | Zn | Zn | Sn | Zn | Sn | N | N | N | N | N | N | N | N | | P2$_1$ |
| Zn | Sn | Sn | Zn | Zn | Sn | Sn | Zn | N | N | N | N | N | N | N | N | | Pca2$_1$ |
| Zn | Sn | Sn | Zn | Sn | Zn | Sn | Zn | N | N | N | N | N | N | N | N | | P2$_1$ |
| Zn | Sn | Sn | Zn | Sn | Zn | Zn | Sn | N | N | N | N | N | N | N | N | | Pna2$_1$ |
| Zn | Sn | Sn | Zn | Sn | Sn | Zn | Zn | N | N | N | N | N | N | N | N | | P1 |
| Zn | Sn | Sn | Sn | Zn | Zn | Sn | Zn | N | N | N | N | N | N | N | N | | P1 |
| Zn | Sn | Sn | Sn | Zn | Zn | Zn | Sn | N | N | N | N | N | N | N | N | | P1 |
| Zn | Sn | Sn | Sn | Zn | Sn | Zn | Zn | N | N | N | N | N | N | N | N | | Pm |
| Zn | Sn | Sn | Sn | Sn | Zn | Zn | Zn | N | N | N | N | N | N | N | N | | Pm |

Table 2A. Basis of the 35 ZnSnN$_2$ 16-atom unit cells and the resultant space group.

Appendix B: Strong coupling condition in the wurtzite-based heterovalent ternary semiconductors

Here, we apply the methods of Holakovsky (1973) to determine the strong coupling condition for the wurtzite-based heterovalent ternaries.

Starting from Eq. 2,

$$\Phi = \Phi_0 + \frac{A}{2}a^2 + \frac{B}{2}b^2 + \frac{C}{4}a^4 - \frac{D}{2}a^2b^2 + \frac{E}{4}b^4 + \frac{F}{6}a^6 - \frac{G}{2}a^4b^2 - \frac{H}{2}a^2b^4 + \frac{I}{6}b^6 \qquad (2)$$

We first minimize in terms of the $a$ order parameter,

$$a(Fa^4 + (C - 2Gb^2)a^2 + A - Db^2 - Hb^4) = 0 \qquad (1A)$$

which has two solutions,

I. $\quad a_I = 0$ (2A)

II. $\quad a_{II}^2 = \frac{-(C-2Gb^2)\pm\sqrt{(C-Gb^2)^2 - 4F(A-Db^2-Hb^4)}}{2F}$ (3A)

Solution I. is valid if $(C - 2Gb^2)^2 < 4F(A - Db^2 - Hb^4)$
Solution II. is valid if $(C - 2Gb^2)^2 > 4F(A - Db^2 - Hb^4)$
At $T_a$, $\qquad (C - 2Gb^2)^2 = 4F(A - Db^2 - Hb^4)$

Since we are considering $T \cong T_a$, we approximate,

$$a_{II}^2 \cong \frac{(2Gb^2 - C)}{2F} \qquad (4A)$$

Inserting 2A and 4A into 1A yields the two distinct solutions which correspond to local minima in $\phi$.

$$\phi_I(a_I) = \Phi_0 + \frac{B}{2}b^2 + \frac{E}{4}b^4 + \frac{I}{6}b^6 \qquad (5A)$$

$$\phi_{II}(a_{II}^2) = \Phi_0 + \left(\frac{C^3}{24F^2} - \frac{AC}{4F}\right) + \frac{B'}{2}b^2 + \frac{E'}{4}b^4 + \frac{I'}{6}b^6 \qquad (6A)$$

where,

$$B' = B + \frac{(2AG + DC)}{2F} - \frac{C^2 G}{2F^2}$$

$$E' = E + \frac{(CH - 2DG)}{F} + \frac{2CG^2}{F^2}$$

$$I' = I - \frac{3HG}{F} - \frac{2G^3}{F^2}$$

To interpret these equations, we consider the behavior of the order parameters as the temperature is lowered.

When $T > T_a$, the displacements associated with both the $a$ and $b$ order parameters are inactive and $a = b = 0$.

At $T = T_a$, the displacements associated with the $a$ order parameter are activated and the coefficients $B$, $E$, and $I$ change to $B'$, $E'$, and $I'$.

Just below $T_a$, the coupling of $a$ to $b$ causes an instability in $b$. The coefficient $B'$ is temperature dependent (3) and it will become negative. Once $B'$ is less than 0, the second transition at $T_b$ is allowed. The nature of a second phase transition at $T_b$ is determined by $E'$.

    A first order phase transition will occur at $T_b$ is $E' < 0$. In the $T_a > T > T_b$ region where $b = 0$, $\phi_b = Bb^2 + Eb^4 + Ib^6 = 0$

$$b^2 = \frac{-E \pm \sqrt{E^2 - 4BI}}{2B}$$

Once $T_b$ is reached, $E$ will transition to $E'$, and the magnitude of $E'$ will determine the nature of the transition. Two types of first order phase transitions can occur after $E$ transitions to $E'$ at $T_a$:

1) The case in which $-2\sqrt{BI} < E' < 0$, is weak coupling and $T_a > T_b$.
2) When $E' < -2\sqrt{BI}$, the temperature regime $T_a > T > T_b$ is forbidden.

Case 2) is the strong coupling case in which, at $T_a = T_b$, the order parameters $a$ and $b$ become finite simultaneously.